\documentclass[12pt]{article}
\usepackage{geometry}                
\usepackage{graphicx}
\usepackage{bm}
\usepackage{amssymb,amsmath,amsthm}
\usepackage{mathrsfs} 
\def\TODAY{23 January 1989}

\title{\bf Traversable wormholes: \\
Some simple examples}
\author{{\Large Matt Visser}\\[10pt]
Theoretical Division\\
         T--8, Mail Stop B--285\\
         Los Alamos National Laboratory\\
         Los Alamos, New Mexico 87545\\[10pt]
Present address: \\
School of Mathematics, Statistics, and Computer Science\\
Victoria University of Wellington, New Zealand\\[5pt]
{\sf \small matt.visser@mcs.vuw.ac.nz}  }
\date{\TODAY;  \LaTeX-ed  \today}                                           

\begin{document}
\maketitle
\def\d{{\mathrm{d}}}
\newcommand{\scri}{\mathscr{I}}
\newcommand{\sun}{\ensuremath{\odot}}
\def\J{{\mathscr{J}}}
\def\sech{{\mathrm{sech}}}
\newtheorem{theorem}{Theorem}
\def\R{\Re}
\def\throat{\curl\Omega}
\def\ie{\emph{i.e.}}
\def\eg{\emph{e.g.}}
\def\curl{\partial}
\begin{abstract}

Building on the work of Morris, Thorne, and Yurtsever, some
particularly simple examples of traversable wormholes are exhibited.  These
examples are notable both because the analysis is not limited to
spherically symmetric cases, and because it is possible to in some sense
minimize the use of exotic matter.  In particular, it is possible for a
traveller to traverse such a wormhole without passing through a region of
exotic matter.  As in previous analyses, the weak energy condition is
violated in these traversable wormholes.

\vskip 10 pt


\vskip 10 pt


\vskip 10 pt

Physical Review {\bf D39} (1989) 3182--3184.

\vskip 10 pt

DOI: 10.1103/PhysRevD.39.3182	

\end{abstract}

\clearpage

\section{Introduction.}
\setcounter{equation}{0}%
 
There has recently occured a major renaissance in wormhole physics.  Most
energy is being focussed on wormholes as possibly significant
contributions to Quantum Gravity, either in the Euclidean or Minkowski
formalisms.  More interesting, I feel, is the analysis of classical
traversable wormholes performed by Morris and Thorne~\cite{MT},
and by Morris, Thorne, and Yurtsever~\cite{MTY}.  These
authors have seriously studied the question of what the properties of a
classical wormhole would have to be in order for it to be traversable by a
human without fatal effects on the traveller.  A major result of their
investigation was that exotic matter, (matter violating the weak energy
condition), was guaranteed to occur at the throat of a traversable
wormhole.
 
Because of the assumed spherical symmetry of their class of models, this
meant that any traveller transiting the wormhole necessarily had to pass
through a region of exotic matter.  The question of how a human body would
interact with such exotic matter was left open.  In this note I shall
construct a particular class of traversable wormholes that are in general
not spherically symmetric.  Though exotic matter is still present in the
region of the throat, it is then possible for a traveller to avoid regions
of exotic material in his/her/its traversal of the wormhole.  Indeed, it is
possible to obtain wormholes and geodesics such that the traveller feels no
forces, tidal or otherwise, during the trip.
 
The major technical change in the analysis is this: whereas Morris and
Thorne~\cite{MT} assumed a spherically symmetric static wormhole, I shall
assume an utrastatic wormhole (\ie, $g_{00}\equiv 1$), with the exotic
matter confined to a thin layer. I shall dispense with the assumption
of spherical symmetry.
 
\section{The Models.}
\setcounter{equation}{0}%
 
The models I have in mind can be very easily described.  Take {\sl two}
copies of flat Minkowski space, and remove from each identical regions of
the form $\Omega \times \R$, where $\Omega$ is a three--dimensional compact
spacelike hypersurface, and $\R$ is a timelike straight line (\eg, the time
axis).  Then identify these two incomplete spacetimes along the timelike
boundaries $\curl\Omega\times\R$.  The resulting spacetime is geodesically
complete, and posesses two asymtotically flat regions (two universes)
connected by a wormhole.  The throat of the wormhole is just the junction
$\throat$ at which the two original Minkowski spaces are identified.  By
construction it is clear that the resulting spacetime is everywhere Riemann
flat except possibly at the throat.  Consequently we know that the stress
energy tensor in this spacetime is concentrated at the throat, with a delta
function singularity there.
 
The situation described above is ready made for the application of the
junction condition formalism.  (Also known as the surface layer formalism.)
Discussions of this formalism may be found in Misner, Thorne, and Wheeler~\cite{MTW}, 
\S21.13, and in a recent article by Blau,
Guendelman, and Guth~\cite{BGG}.  The basic principle is that
since the metric at the throat is continuous but not differentiable, the
connection is discontinuous, and so the Riemann curvature posesses a delta
function singularity.  The strength of this delta function
singularity can be calculated in terms of the second fundamental form on
both sides of the throat.  Because of the particularly simple form of the
geometry, the second fundamental form at the throat is easily calculated~\cite{McConnell}.  In suitable co--ordinates:
\begin{equation} 
K^i{}_j= \left[\begin{matrix}0&0&0\\
                          0&{1\over\rho_1}&0\\
                          0&0&{1\over\rho_2} \end{matrix} \right]_.
\end{equation}
Here $\rho_1$ and $\rho_2$ are the two principal radii of curvature of the
two--dimensional surface $\curl\Omega$.  (A convex surface has positive
radii of curvature, a concave surface has negative radii of curvature.)
Using the symmetry of the wormhole with respect to interchange of the two
flat regions, it is a standard result of the junction formalism that the
Einstein field equations may be cast in terms of the surface
stress--energy tensor $S^i{}_j$ as:
\begin{equation} 
S^i{}_j = -{1\over4\pi G}\left[K^i{}_j - \delta^i{}_j K^k{}_k \right].
\end{equation}
(Factors of $c$ have been suppressed.)  The surface stress energy tensor
may be interpreted in terms of the surface energy density $\sigma$ and
principal surface tensions $\vartheta_{1,2}$:
\begin{equation} 
S^i{}_j  = \left[ \begin{matrix} -\sigma&0&0\\
                           0&-\vartheta_1&0\\
                           0&0&-\vartheta_2 \end{matrix} \right]_.
\end{equation}
Einstein's field equations now yield:
\begin{equation}
\sigma = -{1\over4\pi G}\cdot \left({1\over\rho_1} +
                                      {1\over\rho_2} \right); \qquad
  \vartheta_1 = -{1\over4\pi G}\cdot{1\over\rho_2}; \qquad
  \vartheta_2 = -{1\over4\pi G}\cdot{1\over\rho_1}.
\end{equation}
This implies that in general ($\curl\Omega$ convex) we will be dealing with
{\sl negative} surface energy density and {\sl negative} surface tensions.
 
It is now easy to see how to build a wormhole such that a traveller
encounters no exotic matter. Simply choose $\Omega$ to have one flat face.
On that face the two principal radii of curvature are infinite and the
stress energy is zero. A traveller encountering such a flat face will feel
no tidal forces and see no matter, exotic or otherwise. Such a traveller
will simply be shunted into the other universe.
 
We have just seen that in order for the throat of the wormhole to be
convex, the surface energy density must be negative.  This behaviour may be
rephrased as a violation of the weak energy hypothesis at the throat of the
wormhole.  The violation of weak energy has been noted before in references~\cite{MT} and~\cite{MTY}, wherein very general arguments for this behaviour were
given.  For the traversable wormholes discussed in this Rapid Communication, the
argument simplifies considerably.  Consider a bundle of light rays
impinging on the throat of the wormhole.  A little thought, (following some
geodesics through the wormhole), will convince one that the throat of the
wormhole acts as a ``perfect mirror'', except that the ``reflected'' light
is shunted into the other universe.  This is enough to imply that a convex
portion of $\curl\Omega$ will de--focus the bundle, whereas a concave
portion of $\curl\Omega$ will focus it.  The ``focussing theorem'' for null
geodesics then immediately implies that convex portions of $\curl\Omega$
violate both the weak energy condition and the averaged weak energy
condition.  (See, \eg, reference~\cite{MTW}, exercise 22.14, page 582, or
Hawking and Ellis~\cite{Hawking-Ellis}.)  There are examples, such
as the Casimir vacuum, of states of quantum fields that violate the weak
energy condition; but it is an open question as to whether quantum field
theory ever permits the averaged weak energy condition to be violated~\cite{MT,MTY}.  Until a concrete example of such violation is found in quantum
field theory, the possibility of traverable wormholes such as those studied
here must be viewed with caution.
 
\section{Polyhedral Wormholes.}
\setcounter{equation}{0}%
 
Now that we have discussed non--symmetric solutions in general, it becomes
useful to consider some more special cases.  What I am trying to do here is
to minimize the use of exotic matter as much as possible.  Let the compact
set $\Omega$ be a cube whose edges and corners have been smoothed by
rounding. Then the throat
$\curl\Omega$ consists of six flat planes (the faces), twelve quarter
cylinders (the
edges), and eight octants of a sphere (the corners).  Let the edge of the
cube be of length $L$, and let the radius of curvature of the cylinders and
spheres be $r$, with $r<<L$.  The stress energy tensor on the six faces is
zero.  On the twelve quarter cylinders comprising the edges the
surface stress energy tensor takes the form:
\begin{equation} 
S^i{}_j = {1\over4\pi G r} \left[\begin{matrix}1&0&0\\
                                            0&1&0\\
                                            0&0&0\end{matrix}\right]_.
\end{equation}
This means that each quarter cylinder supports an energy per unit length of
$\mu = \sigma\cdot{2\pi r\over4} = -{1\over4\pi Gr}\cdot{\pi r\over 2} =
-{1\over8G}$, and a tension $T= \vartheta_1\cdot{2\pi r\over4} = \mu$.
Note that $\mu$ and $T$ are well behaved and finite (though negative) as
$r\to0$.  The energy concentrated on each of the eight octants at the
corners is $E=\sigma\cdot{4\pi r^2\over8} = -{1\over2\pi Gr}\cdot{\pi
r^2\over2} = -{r\over4G}$, which tends to zero as $r\to0$.
 
It is therefore safe to take the $r\to0$ limit.  In this limit $\Omega$
becomes an ordinary (sharp cornered) cube.  The stress energy tensor for
the wormhole is then concentrated entirely on the edges of the cube where
$\mu = T = -{1\over8G} = -1.52\times10^{43}\hbox{Joules}/\hbox{metre} =
-{1\over8}\cdot (\hbox{Planck mass}/\hbox{Planck length})$.  [Note that
although the Planck mass and Planck length are quantum concepts, (\ie, they
depend on $\hbar$), the ratio of a Planck mass to a Planck length is
independent of $\hbar$.]  Needless to say, energies and tensions of this
magnitude (let alone {\sl sign}) are well beyond current technological
capabilities.
 
It is interesting to note that the stress energy present at the edges of
the cube is identical to the stress energy tensor of a {\sl negative
tension} classical string.
To see this, write the Nambu--Goto action in the form:
\begin{equation} 
{\mathcal S} = T\int d^2\xi\; d^4x\; \delta^4(x^\mu-X^\mu(\xi))\;
\sqrt{-\det(h_{\alpha\beta})}.
\end{equation}
where $h_{\alpha\beta}(\xi)= \curl_\alpha X^\mu\; \curl_\beta X^\nu\;
g_{\mu\nu}(X^\mu(\xi))$. Varying with respect to the spacetime metric
yields the classical spacetime stress energy tensor
\begin{equation} 
\Theta^{\mu\nu}(x^\rho) =
-T \int d^2\xi\; \delta^4(x^\rho-X^\rho(\xi))\; h^{\alpha\beta}\;
\curl_\alpha X^\mu\; \curl_\beta X^\nu.
\end{equation}
For a classical string stretched along the $x$ axis we may choose the world
sheet coordinates such that $X^\mu(\xi)\equiv X^\mu(\tau,\sigma) =
(\tau,\sigma,0,0)$.
The stress energy tensor is then quickly calculated to be:
\begin{equation}
 \Theta^\mu{}_\nu(t,x,y,z) = -T\; \delta(y)\; \delta(z) \left[
                          \begin{matrix}1&0&0&0\\
                                  0&1&0&0\\
                                  0&0&0&0\\
                                  0&0&0&0\end{matrix} \right]_.
\end{equation}
This is exactly the algebraic form of the stress energy tensor just
obtained for the edges of a cubical wormhole. Note however, that field
theoretic models of strings lead to positive string tensions. No natural
mechanism for generating negative string tension is currently known.
 
Having dealt with cubical wormholes, generalizations are
immediate.  Firstly, note that the length of the edge $L$ nowhere enters
into the calculation.  This implies that any rectangular prism would do
just as well.  The generalization to $\Omega$ being an arbitrary polyhedron
is also straightforward.  Consider an arbitrary polyhedron with edges and
corners smoothed by rounding.  At each edge the geometry is locally that of
two planes joined by a fraction (${\phi\over2\pi}$) of a cylinder.  Here
$\phi$ is the ``bending angle'' at the edge in question.  The local
geometry at each corner is that of some fraction of a sphere.  As before
the energy concentrated on the corners tends to zero as $r\to0$.  The
energy per unit length concentrated on each edge is now $\mu = -{1\over4\pi
Gr}\cdot{\phi r} = -{\phi\over4\pi G}$. As before the limit $r\to0$ is well
behaved, in which case we obtain $\mu=T={-\phi\over4\pi G}$. It is
instructive to compare this with the known geometry of a single infinite
length classical string.  Note that each edge is surrounded by a total of
$2(\pi + \phi)$ radians, ($\pi + \phi$ radians in each universe.)  Thus the
deficit angle ($\varphi$) at each edge is given by $\varphi = -2\cdot\phi$.
In terms of the deficit angle $\mu=T={\varphi\over8\pi G}$, which is the
usual relationship for classical strings.  An edge of the polyhedron is
said to be convex if the bending angle is positive.  If an edge is convex
the tension is negative.  Conversely, if an edge is concave, the tension at
that edge is positive.

\section{Conclusions.}
\setcounter{equation}{0}%
 
In this paper I have exhibited some rather simple examples of traversable
wormholes.  I have been able to avoid the use of spherical symmetry.
Although these wormholes require the presence of exotic matter, it is
possible to exhibit wormholes and geodesics such that the traveller does
not have to encounter the exotic matter directly.  On the other hand, the
required presence of exotic matter, while it is not a cause for panic, is
certainly a cause for concern.  I have attempted to obtain the required
exotic stress energy by considering the Casimir energy associated with
oscillations of a classical string.  So far I have been unable to do so.
 
Open questions include: the stability of such wormholes against
perturbations, and causal constraints on the construction of such
wormholes.  The big open question, naturally, is whether exotic matter is
in fact obtainible in the laboratory.  The theoretical problems are
daunting, and the technological problems seem completely beyond our reach.


\end{document}